\documentclass[%
superscriptaddress,
preprint,
nolongbibliography,
 amsmath,amssymb,
 aps
]{revtex4-2}

\usepackage{graphicx}
\usepackage{dcolumn}
\usepackage{bm}
\usepackage[hidelinks]{hyperref}

\usepackage{braket}

\newcommand{\eq}[1]{\begin{equation}
\begin{aligned}
#1
\end{aligned}
\end{equation}}

\begin{document}

\preprint{}

\title{Ultralight Dark Matter Detection with a Ferromagnet Lattice}

\author{Dongyi Yang}
\affiliation{School of Physics, Peking University, Beijing 100871, China}

\author{Xiao Yang}
\affiliation{Beijing Key Laboratory of Quantum Sensing and Precision Measurement, and Center for Quantum Information Technology, and Institute of Quantum Electronics, Peking University, Beijing 100871, China}

\author{Chenxi Sun}
\email{scx@pku.edu.cn}
\affiliation{Beijing Key Laboratory of Quantum Sensing and Precision Measurement, and Center for Quantum Information Technology, and Institute of Quantum Electronics, Peking University, Beijing 100871, China}

\author{Jianwei Zhang}
\email{james@pku.edu.cn}
\affiliation{School of Physics, Peking University, Beijing 100871, China}




\date{\today}

\begin{abstract}
A levitated ferromagnet provides an exceptionally sensitive probe of ultralight dark matter (ULDM) through measuring weak magnetic-like field signals. We propose a ferromagnet lattice magnetometer that coherently combines multiple levitated ferromagnets to enhance effective sensitivity. By replacing a single ferromagnet with a lattice, we increase the total polarized spin while preserving the intrinsic dynamical response of each constituent ferromagnet. We show that dipole-dipole interactions reshape the collective spectrum, but a spatially uniform magnetic-like signal can still be read out through the coherent mode when mode mixing is kept perturbative by sufficient lattice spacing. We analyze the noise properties of the lattice and demonstrate that collective readout leads to favorable scaling with the number of ferromagnets. Interpreted in terms of axion-electron, dark photon, and axion-photon couplings, our results yield projected sensitivities that exceed existing single-ferromagnet implementations by 6 orders of magnitude.

\end{abstract}

\maketitle
\newpage



Ultralight dark matter (ULDM) is a class of well-motivated dark matter models described by a scalar particle, whose extremely small mass allows it to behave as a coherently oscillating classical field on laboratory timescales \cite{davidj.e.marshAxionCosmology2016,antypas_new_2022}. Through its couplings to standard model particles, ULDM induces spin-dependent interactions that can be described as weak, narrowband, and low-frequency magnetic-like fields, making high-precision magnetometry well-suited for its detection \cite{jackson_kimball_search_2023,cong_spin-dependent_2025}. In this context, spin-based magnetometers with long coherence times are especially powerful, and have enabled a broad experimental program targeting axions, dark photons, and related candidates \cite{fadeev_ferromagnetic_2021,afachSearchTopologicalDefect2021}.

Among various magnetometric platforms, levitated ferromagnets offer a distinctive combination of large intrinsic spin polarization, low mechanical dissipation, and long coherence times \cite{ahrens_levitated_2025,kaliaUltralightDarkMatter2024}. The macroscopic magnetic moment of a single ferromagnet enables coherent rotational dynamics in response to extremely weak magnetic fields, which can be read out with high precision using a superconducting quantum interference device (SQUID). Importantly, levitation isolates the ferromagnet from environmental mechanical noise, allowing its spin dynamics to operate close to fundamental limits. These features make levitated ferromagnets a particularly suitable platform for ULDM searches in the low-frequency regime. A natural strategy to further enhance the detection capability is to increase the total number of polarized spins participating in the measurement. For a single ferromagnet, this can be attempted by enlarging its size. However, practical constraints on levitation and the rapid growth of the moment of inertia limit this approach and degrade the high-frequency response \cite{vinante_levitated_2022}. 

In this letter, we instead propose a qualitatively different route by replacing a single levitated ferromagnet with a lattice of $N$ identical ferromagnets. The lattice increases the total polarized spin and enables a collective readout that improves the effective sensitivity to ULDM-induced magnetic-like fields. Inter-particle magnetic dipole-dipole interactions are the main many-body effect in such a lattice. We describe its response in terms of collective modes, and the separation of the spatially uniform magnetic field and nonuniform modes provides a way of extracting the desired signal from the lattice. As a result, the lattice preserves single-ferromagnet dynamics while achieving a strong enhancement in effective signal-to-noise ratio (SNR).

Beyond this general enhancement, the ferromagnet lattice exhibits a particularly distinctive advantage in the axion-photon coupling channel. Unlike axion-electron or dark-photon interactions, where the ULDM-induced magnetic-like field is determined solely by the dark matter background, the axion-photon signal depends linearly on the ambient electromagnetic field. In our setup, this background field is generated collectively by the ferromagnet lattice itself, leading to an additional enhancement of the effective signal that scales with the number of ferromagnets.


The dynamics of a single levitated ferromagnetic particle are described by its total angular momentum $\mathbf J=\mathbf L+\mathbf S$ and the orientation of its intrinsic spin, $\hat{\mathbf n}=\mathbf S/|\mathbf S|$. In the presence of an external potential $V$, the equations of motion read
\begin{equation}
\partial_t \mathbf J = -\hat{\mathbf n}\times\nabla_{\hat{\mathbf n}}V,
\qquad
\partial_t \hat{\mathbf n}=\boldsymbol\Omega\times\hat{\mathbf n},
\end{equation}
where $\boldsymbol\Omega$ is the angular velocity and $\mathbf L=I\boldsymbol\Omega$ defines the moment of inertia $I$. For small angular deviations about an equilibrium orientation $(\theta_0,\phi_0)$, the equations of motion can be linearized and expressed in terms of angular fluctuations $(\delta\theta,\delta\phi)$. The resulting dynamics are characterized by a susceptibility matrix $\chi(\omega)$, which relates an applied oscillating magnetic field to the angular response of the ferromagnet. When exposed to a weak alternating-current (ac) magnetic field $\mathbf B(t)=\mathbf B_0\cos\omega t$, the response takes the form \cite{kaliaUltralightDarkMatter2024}
\begin{equation}
\binom{\delta\theta}{\delta\phi}
=
-\mu B_0\,
\chi(\omega)
\binom{b_\theta}{b_\phi},
\end{equation}
where $\mu$ is the magnetic moment of the ferromagnet and $(b_\theta,b_\phi)$ specifies the direction of the magnetic field.

The susceptibility matrix encodes oscillatory motions in its diagonal elements and precessional motions in its off-diagonal elements. In the operating regime relevant for magnetometry, the trapping potential is sufficiently strong to dominate the motion of the ferromagnet by oscillations. The detailed form of $\chi(\omega)$ and the derivation of the linearized equations of motion can be found in ref.~\cite{kaliaUltralightDarkMatter2024}, and for our purposes, its simplified expression is
\eq{
\chi(\omega)^{-1}= I\left[-\omega^2\left(\begin{array}{ll}
1 & 0 \\
0 & 1
\end{array}\right)+\omega_\mathrm I\left(\begin{array}{cc}
v_{\theta \theta} & 0 \\
0 & v_{\phi \phi}
\end{array}\right)\right],
\label{eq:chi}
}
where $\omega_\mathrm I = \mu/\gamma_\mathrm e I$ is the Einstein-de Haas frequency, $\gamma_\mathrm e$ is the electron gyromagnetic ratio, and $v_{ij} = (\gamma_{\mathrm{e}}/\mu)\partial_i\partial_j V$, with $i,j = \theta,\phi$.



We now generalize the single-ferromagnet description to a lattice consisting of $N$ identical ferromagnets. The ferromagnets are aligned at fixed positions $\mathbf r_i$, forming a stationary crystal-like lattice configuration. The levitation and spatial separation of multiple ferromagnets could be realized using acoustic standing-wave fields or ultrasonic phased-array techniques, which are capable of generating multiple stable trapping sites and lattice-like configurations for suspended particles \cite{marzo_holographic_2015,chen_acoustic_2019,foresti_acoustophoretic_2013}. An alternative approach is through optical levitation \cite{tseng_search_2025}. Superconducting levitation strongly restricts the vertical extension of the lattice and is thus unfavorable for the lattice configuration.

When extending levitated ferromagnet magnetometry from a single particle to a lattice, magnetic dipole-dipole interactions naturally emerge as the dominant many-body effect. Each ferromagnet produces a magnetic field that acts on the others, so that the response of the array is no longer a simple sum of independent single-particle responses. The interaction introduces collective normal modes of the ferromagnets' angular motion across the lattice. The spatially uniform mode, in which all ferromagnets rotate simultaneously, is the coherent mode that carries the signal from a uniform ULDM-induced magnetic-like field. The remaining modes correspond to nonuniform spatial patterns and are not directly driven by a uniform signal in a periodic lattice.

In this mode representation, the primary effect of the dipole-dipole interaction is to shift the resonance frequencies of the collective modes. For an infinite or periodic lattice, translational symmetry keeps modes with different momenta decoupled, so a spatially uniform magnetic-like field couples only to the coherent mode. The uniform signal can therefore be extracted from the coherent channel even in the presence of interactions, up to the interaction-induced shift of its resonance frequency. In a finite lattice, boundary effects and imperfections break exact translational symmetry and weakly mix the coherent mode with nonuniform modes. This mixing modifies the coherent susceptibility and becomes most important near frequencies where a nonuniform collective mode is close to resonance. In such narrow frequency regions, the response can be strongly reshaped by hybridization with nearby nonuniform modes. Away from these regions, however, the measured response remains well described by the coherent channel.

The strength of these interaction effects is controlled by the dipolar energy scale, which decreases rapidly with the lattice spacing. The most direct and robust strategy is therefore to choose a sufficiently large separation between neighboring ferromagnets, so that the dipole-dipole interaction remains a perturbative correction to the trapping dynamics. This choice trades compactness for theoretical and experimental simplicity. Other approaches, such as dynamical decoupling by external modulation \cite{joseph_decoupling_2025,lang_dynamical-decoupling-based_2015} or optimized lattice geometries that reduce unfavorable dipolar couplings, may further relax the required spacing and enable smaller arrays.

The magnetic-field noise of the ferromagnet lattice can be decomposed into three contributions \cite{clerk_introduction_2010},
\begin{equation}
S_{B} = S^\mathrm{th}_B + S^\mathrm{imp}_B + S^\mathrm{back}_B,
\end{equation}
where $S_{B}$ denotes the power spectral density. The corresponding single-ferromagnet contributions have been analyzed in detail in ref.~\cite{kaliaUltralightDarkMatter2024}; here we focus on their scaling with the number of ferromagnets $N$. The three terms exhibit distinct $N$-dependence. Thermal fluctuations $S^\mathrm{th}_B$, arising from independent environmental perturbations on each particle, are reduced by $1/N$ under collective readout. In contrast, the backaction term $S^\mathrm{back}_B$, originating from SQUID current fluctuations, is fully correlated across the lattice and therefore independent of $N$. The imprecision term $S^\mathrm{imp}_B$ is associated with flux readout fluctuations. Since the collective signal scales linearly with $N$, the equivalent magnetic-field fluctuation is suppressed as $1/N^2$. Neglecting the narrow frequency regions where interaction-induced mode mixing becomes resonantly enhanced, the leading collective scaling of the equivalent magnetic-field noise is
\begin{equation}
S_B =
\frac{1}{N} S^{\rm th}_{B,1}
+ S^{\rm back}_{B,1}
+ \frac{1}{N^2} S^{\rm imp}_{B,1}.
\end{equation}

Although the backaction term is not reduced by increasing $N$, its impact on the total noise can be mitigated through noise rebalancing in the readout. In practice, the magnetic flux from the ferromagnet lattice is collected by a pickup coil and coupled to the SQUID via an effective coupling constant $\eta$. Backaction noise scales as $\eta^2$, while imprecision noise scales as $\eta^{-2}$, allowing their relative contributions to be tuned by adjusting the pickup-coil configuration \cite{kaliaUltralightDarkMatter2024}. Because the imprecision term is suppressed most efficiently by collective operation, it can be partially sacrificed to rebalance the backaction term, such that both contributions can be tuned to be parametrically comparable to the thermal noise floor if possible.

\begin{figure}[ht]
    \centering
    \includegraphics[width=\linewidth]{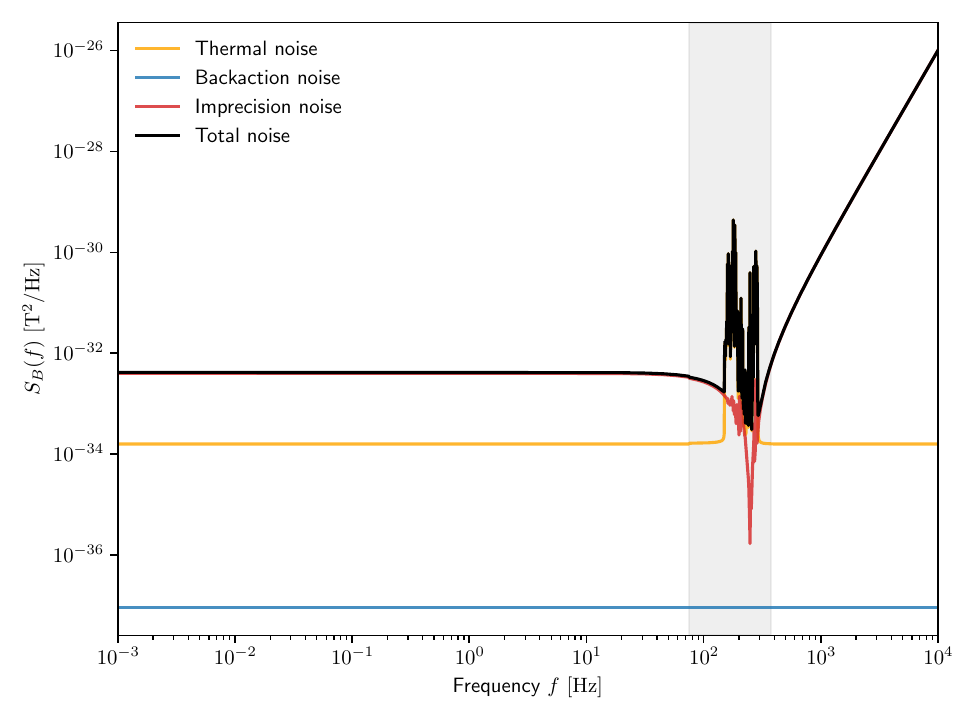}
    \caption{Magnetic-field noise power spectral density $S_{B}(f)$ for the ferromagnet lattice. The total spectrum (black) is shown together with its dominant contributions: thermal, backaction, and imprecision noises. The imprecision term is computed using the frequency-dependent susceptibility $\chi(\omega)$ in Eq.~(\ref{eq:chi}), leading to a pronounced rise at high frequencies. For $v_{\theta\theta} = v_{\phi\phi}$, the system exhibits a single resonance frequency. The gray area indicates the blind frequency range in which near-resonance modes significantly amplify the thermal noise.}
    \label{fig:noise_spectrum}
\end{figure}

Other possible corrections include residual dipole-dipole interactions, finite-size boundary effects, and small variations in ferromagnet properties across the lattice. These effects do not appear as independent stochastic torques. Instead, they modify the collective susceptibility by shifting mode frequencies and by mixing the coherent mode with nonuniform modes. Their impact is therefore frequency dependent. Away from narrow regions where a nonuniform mode becomes nearly resonant with the coherent channel, the correction is perturbative and the collective noise scaling above remains a good approximation. Near such regions, as thermal noise contributes to all lattice modes rather than only the coherent mode, the thermal noise can be significantly amplified, producing frequency bands that should be avoided or treated with a more detailed multimode calculation. Fabrication-induced variations have a similar structure to the boundary effects, as they weakly break the ideal mode separation and introduce mode mixing, but do not introduce a new fundamental noise mechanism. The frequency dependence of the total noise spectrum of a typical parameter setup is shown in FIG. \ref{fig:noise_spectrum}, with detailed parameter setup and calculations shown in ref.~\cite{yang_collective_2026}.


\begin{figure*}[t]
    \begin{tabular}{ccc}
    \includegraphics[width=0.33\linewidth]{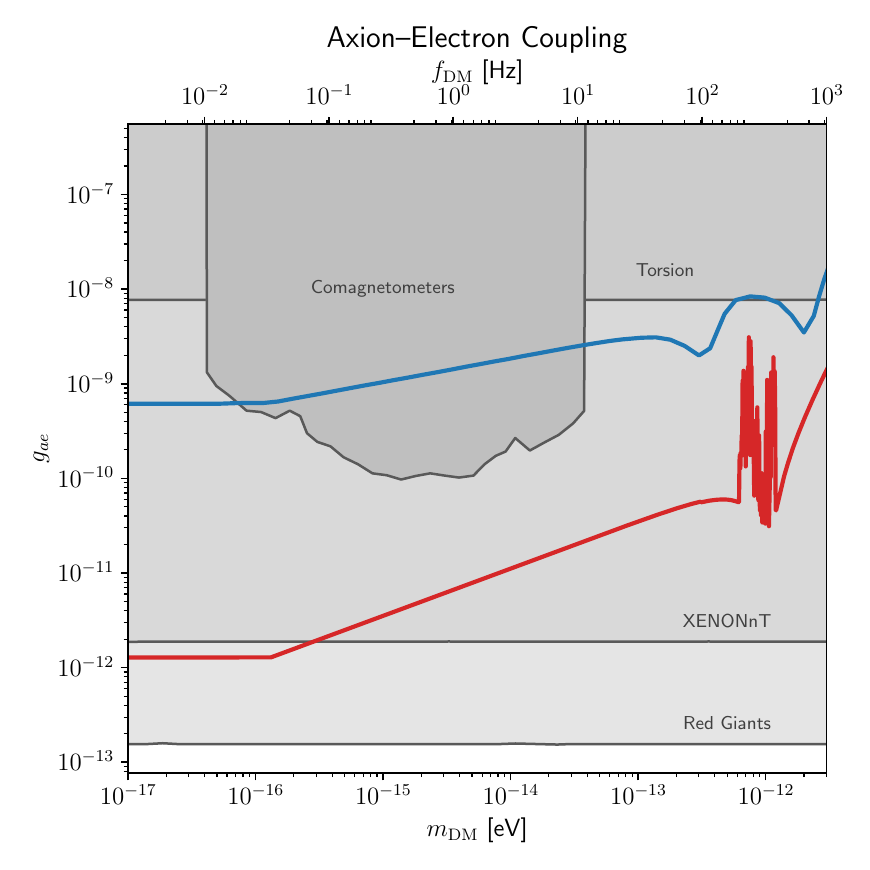} & 
    \includegraphics[width=0.33\linewidth]{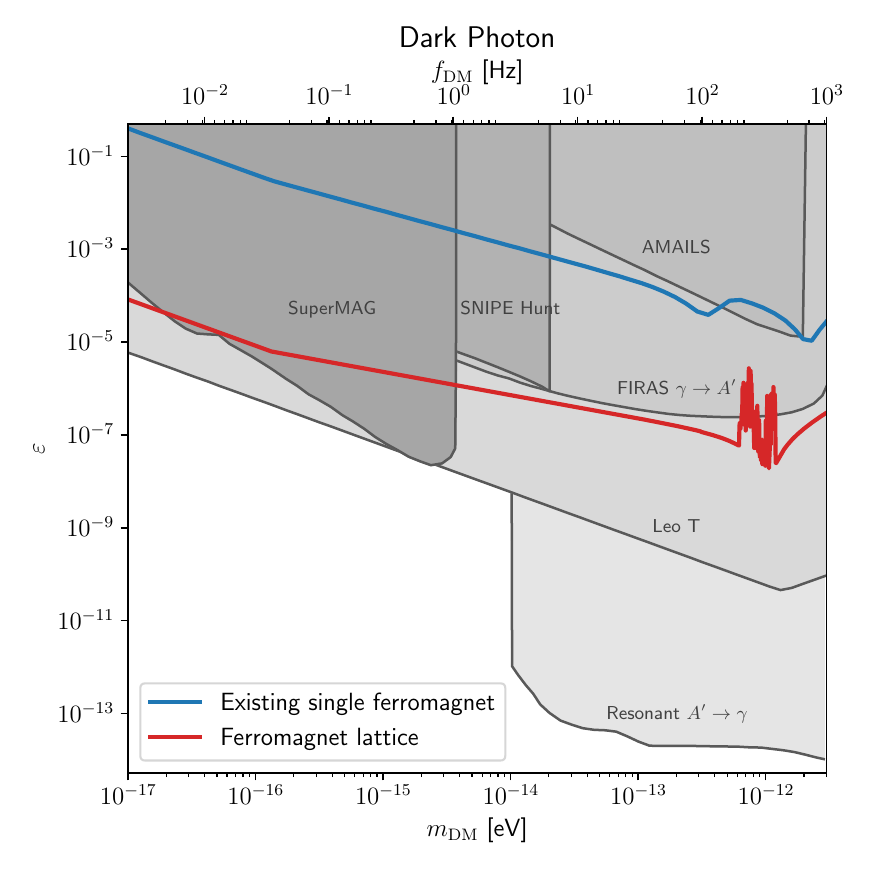} &
    \includegraphics[width=0.33\linewidth]{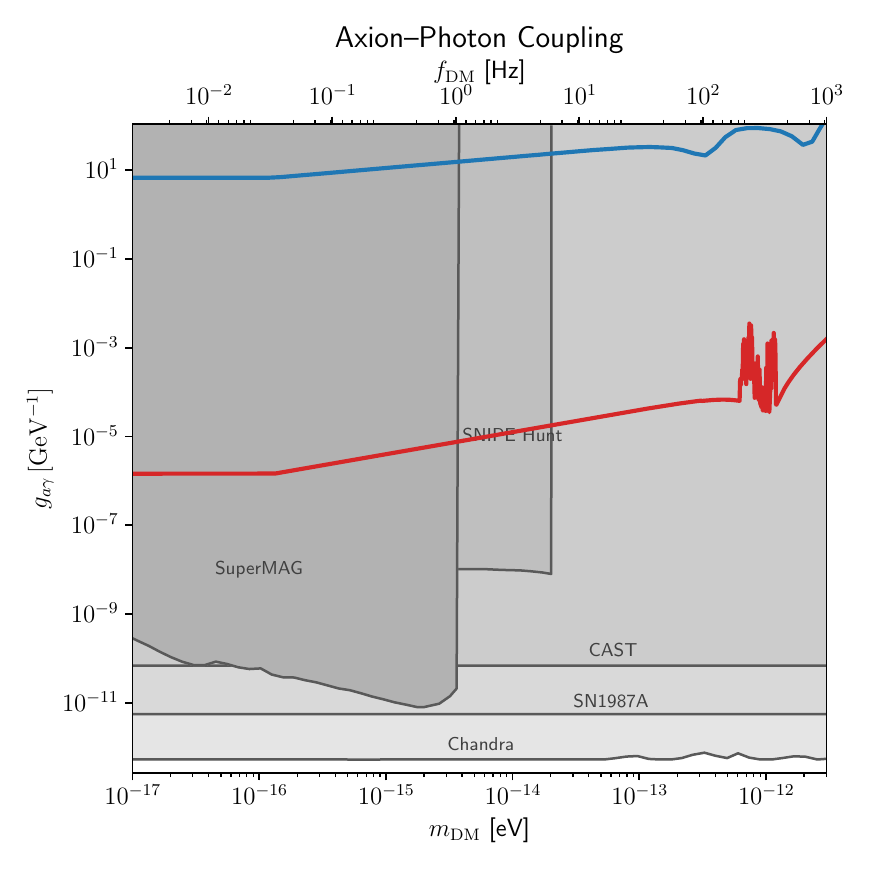}    \\
        (a)&(b)&(c)
    \end{tabular}
    \caption{Projected sensitivity to ultralight dark matter couplings as a function of the dark matter mass $m_\mathrm{DM}$, derived from the projected noise of the proposed ferromagnet lattice magnetometer. The solid red curve shows the constraints obtained from this work. Shaded regions indicate existing constraints from the literature \cite{caputo_dark_2021-1,AxionLimits}. The blue curve denotes the existing experimental constraints from a single ferromagnet magnetometer \cite{kaliaUltralightDarkMatter2024}. In our calculations, we take the size of the shield as $L = 1\ \mathrm{m}$, and use the same parameter setups as the existing single ferromagnet.}
    \label{fig:DMlimits}
\end{figure*}


In the following, we interpret the magnetic-field sensitivity of the proposed ferromagnet lattice magnetometer in the context of ULDM searches. We consider several well-motivated scenarios where ULDM manifests as a coherently oscillating classical field, inducing an effective oscillatory magnetic field in the laboratory. The absence of an experimentally observable signal could allow us to place constraints on the corresponding coupling strengths. We assume a local dark matter energy density $\rho_{\mathrm{DM}}$ and that the dark matter field oscillates at its Compton frequency $\omega_{\mathrm{DM}} = m_{\mathrm{DM}}$, using natural units where $\hbar = c = \mu_0 = 1$ \cite{davidj.e.marshAxionCosmology2016}. The bandwidth of the effective magnetic field is given by its energy dispersion as $\delta f_\mathrm{DM}\equiv 1/t_\mathrm{cor} \sim v_\mathrm{DM}^2 \omega_\mathrm{DM}/2\pi$, resulting in a signal-to-noise ratio (SNR) of \cite{kaliaUltralightDarkMatter2024}
\eq{
    \text{SNR} = \frac{B_\mathrm{DM}^2}{6} \sqrt{\operatorname{Tr}\left[\left(S^{\mathrm{tot}}\right)^{-2}\right] \cdot t_{\mathrm{int}} \cdot \min \left(t_{\mathrm{int}}, t_{\mathrm{coh}}\right)},
}
where $t_\mathrm{int} = 1\ \mathrm{year}$ is the integration time. A criterion of $\text{SNR}\leq 3$ is assumed to set the upper limit.

We first consider an axion-like particle $a$ coupled to electrons through its derivative
\begin{equation}
\mathcal{L}_\mathrm{ae} = \frac{g_\mathrm{ae}}{2m_{\rm e}}\, \partial_\mu a \, \bar{\psi}_{\rm e} \gamma^\mu \gamma^5 \psi_{\rm e},
\end{equation}
which induces a coupling between the axion field gradient and the electron spin. In the nonrelativistic limit, the axion interaction leads to an effective magnetic field given by
\begin{equation}
\mathbf B_\mathrm{ae} = \frac{g_\mathrm{ae}}{\gamma_{\mathrm{e}} m_e}\nabla a.
\end{equation}
For axion dark matter, the axion wind provides a velocity $\mathbf v_\mathrm{DM}\sim 10^{-3}c$, which induces an effective magnetic field oscillating with frequency $\omega_\mathrm{DM}$, reading
\begin{equation}
\mathbf{B}_\mathrm{ae} = \frac{g_\mathrm{ae}}{\gamma_{\mathrm{e}} m_\mathrm e} \mathbf v_\mathrm{DM} \sqrt{2 \rho_\mathrm{DM}} \sin(\omega_\mathrm{DM} t).
\end{equation}
The constraint on the axion-electron coupling constant is shown in FIG. \ref{fig:DMlimits}(a).

Next, we consider dark photon dark matter, characterized by a massive vector field $A'_\mu$ kinetically mixed with the standard model photon through 
\begin{equation}
\mathcal{L}_{A'} = -\varepsilon m_{A'}^2 A_{\mu} A'^{\mu}.
\end{equation}
The background dark photon field generates an effective current
\begin{equation}
J_\mathrm{eff}^\mu = \varepsilon m_{A'}^2 A'^{\mu},
\end{equation}
which oscillates at its Compton frequency. The corresponding effective magnetic field is solved under the boundary condition of the magnetic shield, giving
\begin{equation}
B_{A'} \sim \epsilon \sqrt{2 \rho_{\mathrm{DM}}} m_{A'} L,
\end{equation}
where $L$ is the characteristic length of the shield. The corresponding constraint is shown in FIG. \ref{fig:DMlimits}(b).

Finally, we consider the axion coupling to standard model photons,
\begin{equation}
\mathcal{L}_\mathrm{a\gamma} = -\frac{1}{4} g_\mathrm{a\gamma} a F_{\mu\nu} \tilde{F}^{\mu\nu},
\end{equation}
where $g_\mathrm{a\gamma}$ denotes the axion--photon coupling constant. In the presence of a background electromagnetic field, the axion field induces an oscillatory electromagnetic response that can be described by an effective current
\begin{equation}
J^\mu_\mathrm{eff} = g_\mathrm{a\gamma} \partial_\nu a \tilde{F}^{\mu\nu}.
\end{equation}
This mechanism underlies resonant axion searches in shielded or cavity environments \cite{chaudhuri_radio_2015,ouellet_first_2019,asztalos_squid-based_2010}, where the axion-induced current stimulates electromagnetic modes determined by the boundary conditions of the apparatus. A crucial distinction from the axion-electron and dark-photon cases considered above is that the induced signal depends linearly on the ambient electromagnetic field. The experimental configuration therefore participates directly in the signal-generation process rather than acting as a passive probe. In our setup, the dominant background field is generated by the ferromagnets themselves. The magnetic shield imposes boundary conditions that define a discrete set of cavity modes, as analyzed in detail in ref.~\cite{higgins_maglev_2024}. Since both the lattice size and the minimum distance to the shield are much smaller than the characteristic shield dimension $L$, all ferromagnets couple identically to these low-frequency modes. These modes therefore mediate a collective electromagnetic response, in which the contributions from individual ferromagnets add coherently. Because the axion-induced signal is dominated by these low-frequency cavity modes, the resulting magnetic-like field experienced by the ferromagnets scales linearly with the number of ferromagnets, reading
\eq{
B_\mathrm{a\gamma}\sim N B_\mathrm{a\gamma}^{(1)} 
= \mathcal O(0.1)\, N \, g_\mathrm{a\gamma}\, 
\sqrt{2\rho_{\mathrm{DM}}}\frac{\mu}{L^2},
}
where the single-ferromagnet estimate follows 
refs.~\cite{kaliaUltralightDarkMatter2024,higgins_maglev_2024}. The numerical prefactor depends on the detailed cavity geometry but remains $\mathcal O(0.1)$ for generic shield configurations.

This linear enhancement is qualitatively different from the scaling obtained for axion-electron or dark-photon couplings, where collective operation improves only the effective sensitivity through signal averaging. In the axion-photon channel, the ferromagnet lattice enhances the signal generation mechanism itself, leading to a genuinely nontrivial amplification of the axion-induced magnetic response. The resulting projected sensitivity is shown in FIG.~\ref{fig:DMlimits}(c).

 In conclusion, we propose and analyze a ferromagnet lattice magnetometer as a new platform for ultra-sensitive detection of weak effective magnetic fields induced by ULDM. By replacing a single ferromagnet with a lattice of identical ferromagnets, the total polarized spin is increased, thereby amplifying the signal strength without introducing dynamical complications associated with scaling up a single ferromagnet. We show that magnetic dipole-dipole interactions can be incorporated through a collective-mode description. A spatially uniform magnetic-like field couples to the coherent mode of the lattice, while the nonuniform modes mainly enter through interaction-induced frequency shifts and weak finite-size mixing. By choosing a sufficiently large lattice spacing, these corrections can be kept perturbative over most of the relevant frequency range, allowing the coherent channel to retain the favorable collective scaling. We perform a noise analysis in this coherent-channel regime and demonstrate how each noise component scales with the number of ferromagnets. We perform a detailed analysis of the noise properties of the driven lattice and demonstrate how each noise component scales with the number of ferromagnets. Although backaction noise remains correlated across the lattice, we show that its impact can be mitigated through appropriate tuning of the pickup-coil coupling, allowing the total noise to be rebalanced near the thermal noise floor. As a result, compared to existing single ferromagnetic implementations, the lattice configuration achieves $N$ times higher effective sensitivity. We interpret the projected sensitivity in the context of ULDM searches, focusing on axion-electron, dark photon, and axion-photon couplings. While collective operation improves sensitivity in all cases, we find that the axion-photon channel exhibits a qualitatively distinct enhancement, as the electromagnetic background generated by the lattice itself coherently amplifies the axion-induced signal. Our results establish the ferromagnet lattice magnetometer as a promising platform for future searches for ultralight dark matter, with clear pathways for further improvement through lower-noise SQUID readout, optimized pickup-coil geometries, larger arrays, and additional methods for reducing residual dipolar mixing.

\begin{acknowledgments}
    \textit{Acknowledgments}--This work is supported by the National Natural Science Foundation of China (Grant No. 62505007).
\end{acknowledgments}

\bibliographystyle{apsrev4-2_mod}
\bibliography{bib_deduplicated}

\end{document}